\documentclass[aps,prl,superscriptaddress,reprint]{revtex4-1}

\usepackage[usenames, dvipsnames]{color}

\usepackage{graphicx}
\usepackage{xcolor}
\usepackage{rotating}
\usepackage{bm,amsmath,amssymb}
\usepackage[utf8]{inputenc}
\usepackage{footmisc}

\DeclareMathAlphabet{\mathpzc}{OT1}{pzc}{m}{it}

\usepackage{setspace}

\usepackage{slashed}

\newcommand{\nb}[1]{\color{blue}}

\newcommand{\para}{{\parallel}}

\usepackage[
  pagebackref=false,
  colorlinks=true,
  linkcolor=blue,
  urlcolor=blue,
  filecolor=black,
  citecolor=red,
  pdfstartview=FitV,
  pdftitle={},
  pdfauthor={},
  pdfsubject={},
  pdfkeywords={},
  pdfpagemode=None,
  bookmarksopen=true
]{hyperref}

\usepackage{graphicx}
\graphicspath{ {./images/} }

\usepackage{comment}

\usepackage[normalem]{ulem}
\usepackage{amsmath}
\usepackage{enumerate}
\usepackage{amsfonts}
\usepackage{epsfig}
\usepackage{mathbbol}

\def\Tr{\mathop{\rm Tr}}

\def\Im{\mathop{\rm Im} }
\def\Re{\mathop{\rm Re} }

\newcommand\half{{\ensuremath{\frac{1}{2}}}}

\newcommand\ket[1]{\ensuremath{\lvert{#1}\rangle}}
\newcommand\bra[1]{\ensuremath{\langle{#1}\rvert}}

\newcommand{\be}{
\begin{equation}}
  \newcommand{\ee}{
\end{equation}}
\newcommand{\bea}{
\begin{eqnarray}}
  \newcommand{\eea}{
\end{eqnarray}}
\newcommand{\bega}{
\begin{gather}}
  \newcommand{\eega}{
\end{gather}}

\newcommand{\bi}{
\begin{itemize}}
    \newcommand{\ei}{
\end{itemize}}
\newcommand{\ben}{
\begin{enumerate}}
    \newcommand{\een}{
\end{enumerate}}
\newcommand{\bca}{
\begin{cases}}
  \newcommand{\eca}{
\end{cases}}
\newcommand{\bln}{
\begin{align}}
  \newcommand{\eln}{
\end{align}}
\newcommand{\bst}{
\begin{split}}
  \newcommand{\est}{
\end{split}}
\def\ie{
  \begin{equation}
  \begin{aligned}}
    \def\fe{
    \end{aligned}
\end{equation}}
\newcommand{\bma}{\le(
  \begin{matrix}}
    \newcommand{\ema}{
\end{matrix}\ri)}
\newcommand{\bwt}{
\begin{widetext}}
  \newcommand{\ewt}{
\end{widetext}}

\newcommand\ep{\epsilon}
\newcommand\sig{\sigma}

\newcommand\om{\omega}

\newcommand\da{{\dagger}}

\newcommand\ra{{\rightarrow}}

\def\le{\left}
\def\ri{\right}

\newcommand\sE{{\ensuremath{{\mathcal E}}}}

\newcommand\sM{{\ensuremath{{\mathcal M}}}}

\newcommand\vx{{\bm x}}
\newcommand\vk{{\bm k}}

\DeclareMathAlphabet{\pazocal}{OMS}{zplm}{m}{n}

\IfFileExists{latexml.sty}{\usepackage{latexml}}{\newif\iflatexml}  % true only in arXiv's HTML (LaTeXML) build
\newcommand{\vq}{{\bm q}}                  % momentum transfer
\newcommand{\vK}{{\bm K}}                  % k_i + k_f
\newcommand{\vP}{{\bm P}}                  % spin polarization
\newcommand{\Mh}{{\hat{\bm M}}}            % direction of the ordered spins
\newcommand{\qhp}{{\hat{\bm q}_\perp}}     % transverse momentum-transfer direction
\newcommand{\nh}{{\hat{\bm n}}}            % scattering-plane normal
\newcommand{\bL}{{\bm L_{\sE}}}            % spin angular momentum of the mode
\newcommand{\bsE}{{\bm{\mathcal E}}}       % target spin matrix element
\newcommand{\eS}{{\epsilon_X}}             % exchange-to-dipole amplitude ratio

\begin{document}

% \title{Electron exchange in magnon energy-loss spectroscopy\\ is a which-path measurement}
\title{Electron exchange matters in electron spectroscopy}
\author{ Michael J. Landry}
\email{mjlandry@mit.edu}
\affiliation{Quantum Measurement Group, MIT, Cambridge, MA 02139, USA}
\affiliation{Department of Nuclear Science and Engineering, MIT, Cambridge, MA 02139, USA}
\affiliation{Department of Physics, MIT, Cambridge, MA 02139, USA}

\author{Chuhang Liu}
\affiliation{Condensed Matter Physics and Materials Science Department, Brookhaven National Laboratory, Upton, NY 11973, USA}
\affiliation{Department of Physics and Astronomy, Stony Brook University, Stony Brook, NY 11794, USA}

\author{Ashley S. Dale}
\affiliation{Quantum Measurement Group, MIT, Cambridge, MA 02139, USA}
\affiliation{Laboratory for Information and Decision Systems, MIT, Cambridge, MA 02139, USA}

\author{L. Dagny Davis}
\affiliation{Quantum Measurement Group, MIT, Cambridge, MA 02139, USA}
\affiliation{Department of Nuclear Science and Engineering, MIT, Cambridge, MA 02139, USA}

\author{Griffin M. Haisman}
\affiliation{Quantum Measurement Group, MIT, Cambridge, MA 02139, USA}
\affiliation{Department of Electrical Engineering and Computer Science, MIT, Cambridge, MA 02139, USA}

\author{Lijun Wu}
\affiliation{Condensed Matter Physics and Materials Science Department, Brookhaven National Laboratory, Upton, NY 11973, USA}

\author{Ray F. Egerton}
\affiliation{Physics Department, University of Alberta, Edmonton, AB T6G 2E1, Canada}

\author{Yimei Zhu}
\affiliation{Condensed Matter Physics and Materials Science Department, Brookhaven National Laboratory, Upton, NY 11973, USA}

\author{Mingda Li}
\email{mingda@mit.edu}
\affiliation{Quantum Measurement Group, MIT, Cambridge, MA 02139, USA}
\affiliation{Department of Nuclear Science and Engineering, MIT, Cambridge, MA 02139, USA}

\begin{abstract}
Electron energy-loss spectroscopy (EELS) conventionally considers classical Coulomb interactions between probe and material electrons, overlooking their fermionic exchange at relativistic energies. Here, we develop a relativistic theory of magnon EELS incorporating both magnetic dipole and electron exchange interactions. We show that the two channels leave distinguishable electron spin states, suppressing their interference in conventional EELS. However, spin-resolved detection erases this which-path information, revealing first-order interference despite the weak exchange amplitude. Remarkably, the same interference produces a left–right intensity asymmetry with a spin-polarized incident beam, without requiring outgoing-spin analysis. We demonstrate that this signal directly probes magnon angular momentum and distinguishes excitations of opposite handedness, opening a route to nanoscale measurements of chiral magnetic excitations.

\end{abstract}

\maketitle
{In condensed matter systems, electrons, nuclei and their interactions determine the properties of materials. It is therefore natural to use electrons as probes to study those properties as they interact strongly with the material~\cite{egerton2011}. As charged particles, electrons can be accelerated to relativistic speeds, with kinetic energy of tens to hundreds of keV or higher. The short de Broglie wavelength of relativistic electrons, together with the advances of electron optics in aberration correction, allow electron beam to be focused to atomic scale in a modern transmission electron microscope (TEM), enabling atomic-scale structural characterization through electron imaging and diffraction~\cite{haider1998,batson2002}. Besides atomic structure, TEM can also probe electromagnetic field distributions by measuring phase shifts of the electron wave~\cite{mccartney2007,shibata2012,volkov2002}.}

{To study the key questions in condensed matter physics, such as unconventional superconductivity, topological edge state, quantum spin liquid and altermagnetism, it is also necessary to understand the quasiparticle excitations and energy dispersions, which can not be accessed directly by structural characterization~\cite{christianson2008,bednorz1986,fong1999,drozdov2014,anderson1973,fazekas1974,han2012,LiChen2020, smejkal2022,Chou2024,jost2026}. To this end, electron energy-loss spectroscopy (EELS) probes the energy aspect of the material responses through inelastic electron scattering~\cite{egerton2011}. EELS can measure the electronic transitions, as well as the collective excitations such as plasmons, phonons and magnons~\cite{egerton2011,krivanek2014,kepaptsoglou2025}. EELS can also be used to study magnetism, for example, it can measure static magnetic moments via electron magnetic circular dichroism (EMCD)~\cite{schattschneider2006}. Recent development of monochromated EELS in scanning transmission electron microscopy (STEM) achieved an energy resolution below $\sim$3~meV, enabling the study of low-energy excitations in nanoscale structures~\cite{mm2023nion}.}

{Existing theories of EELS in electron microscopy consider only the energy loss by direct Coulomb interaction and magnetic dipole interaction~\cite{lyon2021,mendis2021,mendis2022,castellanos2023,nascimento2024,nascimento2025,castellanos2025}. Quantum mechanically, these electromagnetic interactions can be described as the exchange of virtual photons between the beam electron and the specimen~\cite{iatrakis2026a,iatrakis2026b}. Beyond these interactions, the indistinguishability of the beam electron and specimen electrons as fermions allows the beam electron to replace a specimen electron, which is ejected with reversed spin due to the Pauli exclusion principle. As a result, the scattering amplitudes also contains the exchange channel, which dominates at low-energy electron spectroscopy~\cite{hopster1984,kirschner1984,kirschner1985,plihal1999,vollmer2003,zakeri2014}. However, the exchange channel is usually omitted in relativistic EELS~\cite{schattschneider2005,sorini2008,iatrakis2026a,iatrakis2026b}, since its amplitude is suppressed at high beam energies~\cite{ochkur1964,rudge1968,porter1994}. At typical electron microscopy energies ranging from 30 to 300~keV, the magnitude of the exchange amplitude relative to the dipole amplitude, as well as the interference between the two channels, has not been calculated.}

{In this Letter, we show that the exchange channel can play an important role in relativistic EELS. Using magnon excitation as an example, we compute both the dipole and exchange channels in one unified relativistic framework~\cite{companion}. For an unpolarized incident beam with outgoing spins indistinguishable, exchange contributes a second-order correction in the exchange-to-dipole amplitude ratio. However, spin-resolved detection reveals a first-order transverse polarization arising from interference between the two channels. This interference can also be detected as an intensity asymmetry using a transversely polarized incident beam without outgoing-spin analysis, providing an alternative route to measuring magnon handedness. We further simulate spin-resolved EELS spectra of model materials, demonstrating the spectral signatures of magnon modes with opposite handedness. These results establish the importance of electron exchange in relativistic EELS and provide a route to probing the chirality of magnetic excitations.}

\begin{figure}[t]
  \includegraphics[width=\columnwidth]{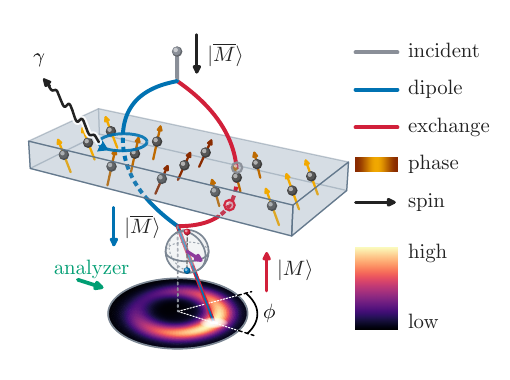}
  \caption{One electron with two scattering amplitudes, $\sM=a\,\mathbb{1}+\bm b\cdot\bm\sig$, not two trajectories. Dipole (blue, used by previous theories): the beam electron's magnetic field (ring) torques the film's spins; it exchanges a virtual photon ($\gamma$, wavy line) with the magnetization current. The beam electron's spin is unchanged. Exchange (red): the beam electron is captured, and a film electron is ejected with reversed spin. The objective-lens field sets the ordered spins against the beam, $\Mh\parallel-\vk_i$. $\ket{\overline{M}}$ is spin against $\Mh$ (along the beam), the incident minority spin, and $\ket{M}$ is spin along $\Mh$. The final spin records the path. $\ket{\overline{M}}$ (dipole) and $\ket{M}$ (exchange) are orthogonal, so a spin-blind readout shows no cross term and a uniform detector ring. Analyzing the spin along a transverse axis (green arrow) erases the record and produces $P_\perp$, and a transversely polarized source produces a $\cos\phi$ detector modulation (exaggerated). Dashed segments lie inside or behind the film. The detector lies below the film, and the color scale gives detector intensity.}
  \label{fig1}
\end{figure}

Let the beam electron have mass $m$, kinetic energy $T$, total energy $E=m+T$, and momentum $\vk_i$ before and $\vk_f$ after scattering, in units $\hbar=c=1$. It transfers momentum $\vq=\vk_i-\vk_f$ and energy $\om$ to the film. Magnon energies are millielectronvolts, and momentum transfers are of order an inverse \aa ngstr\"om, so $\om\ll q\ll k\equiv|\vk_i|$ and the scattering angle is small. We take the film to be a ferromagnet with spin $S$ per site, $n$ sites per unit volume, and all spins ordered along a unit vector $\Mh$~\footnote{$\Mh$ is the direction of the ordered electron \emph{spins}. The electron's magnetic moment is antiparallel to its spin, so the magnetization of the film points along $-\Mh$. This affects only the sign of the predicted polarization.}. A magnon of wavevector $\vq$ is the plane-wave state $\ket{\vq}$ with the total spin lowered by one unit~\cite{holstein1940}. The film enters both channels through the electron spin density $\bm s(\vx)$ (twice the spin-angular-momentum density), whose Fourier component $\bm s_{\vq}=\int d^3x\,e^{i\vq\cdot\vx}\bm s(\vx)$ connects the ground state to the magnon,
\be
\bra{\vq}\bm s_{\vq}\ket{0}=\sqrt{2SN}\,F(q)\,\bm\ep,\qquad
\bm\ep=\hat{\bm e}_1+i\hat{\bm e}_2 ,
\label{eq:eps}
\ee
where $N$ is the number of sites, $F(q)$ is the magnetic form factor, and $(\hat{\bm e}_1,\hat{\bm e}_2,\Mh)$ is a right-handed triad. The complex vector $\bm\ep$ is the circular polarization of the spins precessing about $\Mh$. It is the only property of the film that enters the results below.

The amplitude for creating this magnon is a $2\times2$ matrix in the spin of the beam electron. To leading order in the scattering angle,
\be
\sM=a\,\mathbb{1}+\bm b\cdot\bm\sig ,
\label{eq:M}
\ee
with $\bm\sig$ the Pauli matrices. The dipole channel gives the spin-independent part,
\be
a=-\frac{ie^2}{2m}\,\frac{(\vK\times\vq)\cdot\bm\ep}{q^2},\qquad \vK\equiv\vk_i+\vk_f ,
\label{eq:a}
\ee
where we drop the common factor $\sqrt{2SN}F(q)$ of Eq.~\eqref{eq:eps} throughout. Up to a phase, $(\vK\times\vq)/2mq^2$ is the Fourier component at wavevector $\vq$ of the Biot--Savart field of the moving charge, projected on the precessing spin. The exchange channel gives the spin-dependent part,
\be
\bm b=e^2\,\Pi_X\,m\,\bm\ep,\qquad \Pi_X=\frac{1}{2m(E-m)}=\frac{1}{2mT} .
\label{eq:b}
\ee
Here $\Pi_X$ is the propagator of the virtual photon that carries the beam electron's momentum into the film. This photon absorbs nearly all of $\vk_i$, so it is far off shell, with virtuality $Q^2\simeq-2mT$, and the exchange interaction acts as a contact interaction independent of $q$. Nonrelativistically $2mT=k^2$ and $\Pi_X=1/k^2$ is the Ochkur kernel~\cite{ochkur1964}. The relativistic form holds at all $T$. The component of $\bm b$ along the beam carries $E$ in place of $m$, and it does not enter here because $\bm\ep\perp\Mh$ and we take $\Mh$ along the beam. The dipole amplitude $a$ multiplies the unit matrix because the Biot--Savart field couples to the beam's orbital motion. The exchange amplitude multiplies $\bm\sig\cdot\bm\ep$ because exchange transfers the film's change of spin to the outgoing electron. All results below agree with an exact evaluation using Dirac spinors, and with exact M\o ller scattering off a free electron with a spin flip, to better than $10^{-4}$ at 100~keV~\cite{companion}.

The ratio of the two amplitudes, with $\bm b=b\,\bm\ep$,
\be
\eS\equiv\frac{|b|}{|a|}=\frac{qm}{2Tk},
\label{eq:epsS}
\ee
is the small parameter of the problem. For $q=1$~\AA$^{-1}$ it is $9.5\times10^{-2}$ at 30~keV and $1.5\times10^{-2}$ at 100~keV. It grows linearly with $q$ because the Biot--Savart field falls as $1/q$ and the contact interaction does not. It falls with beam energy as $T^{-3/2}$ nonrelativistically and steepens toward $T^{-2}$ as $T$ approaches $m$.

A conventional experiment measures the intensity with an unpolarized beam and no spin analysis of the scattered electrons. Averaging over the incident spin and summing over the final spin, with $|\bm\ep|^2=\bm\ep^*\cdot\bm\ep=2$, gives
\be
\half\Tr[\sM\sM^\da]=|a|^2+|\bm b|^2=|a|^2\,(1+2\eS^2).
\label{eq:intensity}
\ee

\begin{figure}[t]
  \includegraphics[width=\columnwidth]{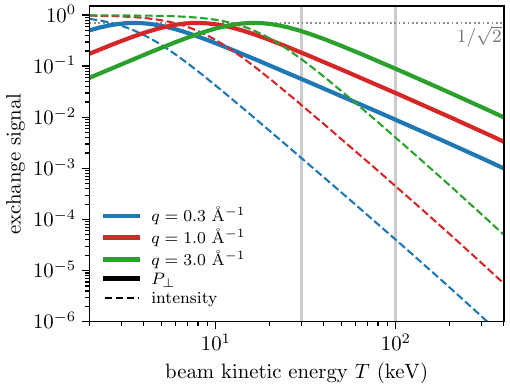}
  \caption{Exchange signal versus beam energy for three momentum transfers. Solid lines: transverse polarization ($P_\perp$), Eq.~\eqref{eq:Pperp}, first order in $\eS$, with maximum $1/\sqrt2$ where the two channels are equal. Dashed lines: fraction of the unpolarized intensity due to exchange, $2\eS^2/(1+2\eS^2)$ from Eq.~\eqref{eq:intensity}, second order in $\eS$. Vertical lines mark 30 and 100~keV.}
  \label{fig2}
\end{figure}

The cross term, which would be first order in $\eS$, is proportional to $\Tr[\bm\sig]$ and vanishes. This holds for any magnet: the exchange amplitude always carries one power of $\bm\sig$, because the exchange interaction is linear in the film's spin density, and the dipole amplitude carries none. Exchange therefore enters the intensity only at $2\eS^2$. For $q=1$~\AA$^{-1}$ this is $2\%$ at 30~keV and $4.5\times10^{-4}$ at 100~keV, and it reaches unity only at $T_\times\simeq(q^2m/4)^{1/3}\approx8$~keV (cf. Fig.~\ref{fig2}, dashed). The dipole-only theories of Refs.~\cite{mendis2022,nascimento2025} are complete for unpolarized microscope experiments at $q\leq1$~\AA$^{-1}$ to $2\%$ at 30~keV and to better than $10^{-3}$ above 100~keV.

The beam electron's spin records which channel acted. The dipole channel leaves the spin unchanged. The exchange channel reverses the spin, because the emerging electron is the ejected one (Fig.~\ref{fig1}). The two final spin states are orthogonal and do not interfere. This is a which-path measurement in the interferometric sense~\cite{scully1982,englert1996}, with the population of each final spin state equal to the sum of the two path probabilities.

The record is stored in a two-state system measurable in any basis, so that measuring the beam electron's spin along an axis transverse to $\Mh$ erases the record and restores the interference. For an incident electron with spin down along $\Mh$, the final state is $a\ket{\downarrow}+2b\ket{\uparrow}$, since $\bm\sig\cdot\bm\ep$ is twice the raising operator. Its populations, $|a|^2$ and $4|b|^2$, are the two path probabilities. A transverse spin measurement reads this final state's coherence between $\uparrow$ and $\downarrow$, which is $2ab^*$, linear in the exchange amplitude. Averaging over the incident spin, the polarization vector of the scattered electrons is
\be
\vP_f=\frac{\Tr[\bm\sig\,\sM\sM^\da]}{\Tr[\sM\sM^\da]}
=\frac{2\,\Re(a\,\bm b^*)+2|b|^2\,\Mh}{|a|^2+2|b|^2}.
\label{eq:Pf}
\ee
The component along $\Mh$ is second order, so that all exchange-scattered electrons emerge with spin along $\Mh$, while the transverse component is first order. In electron--atom collisions the analogous interference of direct and exchange amplitudes appears in the polarization of the scattered electrons~\cite{kessler1985,hanne1983}, but it has not been considered for magnons. With $\Mh$ along the beam, $\Mh=+\hat{\vk}_i$, $\vK\times\vq$ is transverse and
\be
\vP_\perp=\frac{2\eS}{1+2\eS^2}\;\qhp ,
\label{eq:Pperp}
\ee
where $\qhp$ is the direction of the transverse momentum transfer; $\vP_\perp$ reverses for spins against the beam. The scattered electrons are polarized in the scattering plane, opposite to the direction of deflection. For $q=1$~\AA$^{-1}$, the polarization is $19\%$ at 30~keV and $3\%$ at 100~keV; it reaches $1/\sqrt2$ at $T_\times$ (Fig.~\ref{fig2}, solid). At 100~keV the polarization exceeds the intensity effect $2\eS^2$ of Eq.~\eqref{eq:intensity} by a factor of 66. Three corrections to Eq.~\eqref{eq:Pperp} are each below $1\%$ of $P_\perp$ above 30~keV for a spin-only magnet~\cite{companion}: the beam's own spin-magnetization current with relative order $q/2k$, the finite range of the exchange interaction, and the orbital fraction of the moment.

For a beam of polarization $\vP_b$, the intensity with final spins summed is $|a|^2+|\bm b|^2+\vP_b\cdot[\,2\Re(a\,\bm b^*)-2|b|^2\Mh\,]$, and the transverse term, $2\,\mathrm{Re}(a\, \bm b^*)$, equals that of Eq.~\eqref{eq:Pf}. The longitudinal term changes sign because, by the Pauli principle, incident electrons with spin along $\Mh$ cannot exchange. With the beam polarized transversely, the magnon intensity around the optic axis is
\be
I(\phi)=\bar I\,\big[1+P_b\,P_\perp\cos\phi\big],
\label{eq:cosphi}
\ee
where $\phi$ is the azimuth of the detected electron measured from $\vP_b$, for spins against the beam as in Fig.~\ref{fig1}. For spins along the beam the modulation changes sign, because its analyzing power is the same vector $2\Re(a\,\bm b^*)$ that gives $\vP_\perp$, and that vector reverses with $\Mh$. The signature is a left--right asymmetry in a momentum-resolved detector that reverses with the beam polarization and with $\Mh$.

Polarizing the beam antiparallel to $\Mh$ and analyzing the scattered spin removes the Pauli-blocked half of the beam, which contributes only dipole background, and raises the polarization to about $4\eS/(1+4\eS^2)$. This doubles the polarization and quadruples the figure of merit $P^2$, but a spin analyzer has an efficiency of only $\sim10^{-4}$~\cite{gay1992}, so the measurement with a polarized source alone is faster.

The ferromagnetic result is a special case. For any spin excitation $\ket{n}$ of any magnet, collinear or not, let $\bsE=\bra{n}\bm s_{\vq}\ket{0}$ be the complex spin matrix element that creates it, with $\bm\ep$ the ferromagnetic case. The amplitudes remain $a\propto(\vK\times\vq)\cdot\bsE$ and $\bm b\propto\bsE$. Since $\vK\times\vq$ is real,
\be
2\Re(a\,\bm b^*)=\frac{e^4\Pi_X}{m\,q^2}\,\mathsf{D}\big[(\vK\times\vq)\times\bL\big],
\quad
\bL\equiv-\frac{i}{2}\,\bsE^*\!\times\bsE ,
\label{eq:law}
\ee
where $\mathsf{D}=\text{diag}(m,m,E)$ in the beam frame is the relativistic vertex factor noted below Eq.~\eqref{eq:b}. The vector $\bL$ is the spin angular momentum carried by the mode. It equals $\Mh$ for a ferromagnetic magnon, $\pm\hat{\bm N}$ for the two modes of a uniaxial antiferromagnet with N\'eel vector $\hat{\bm N}$, and zero for a linearly polarized mode. The first-order signal therefore measures magnon angular momentum, with three consequences: (i) Modes with no angular momentum give no first-order signal in the intensity for any beam polarization, nor in the polarization of an unpolarized beam. Examples are the breathing mode of a skyrmion lattice, the $\Gamma$-point mode of a spin spiral~\cite{dossantos2018}, and the easy-plane modes of NiO. For real $\bsE$ the amplitude $a$ is imaginary and $\bm b$ is real, so the two paths are $90^\circ$ out of phase, and their cross term survives only as a rotation of the spin of a polarized beam, $\propto\Im(a\,\bm b^*)\times\vP_b$, which needs both a polarized source and a spin analyzer. (ii) The two modes of an unsplit antiferromagnet cancel. A field that splits them turns the signal on, with opposite signs in the two branches. (iii) $\vK\times\vq$ lies along the normal $\nh$ to the scattering plane, so Eq.~\eqref{eq:law} lies in that plane. A transverse polarization along $\qhp$ measures the component of $\bL$ along the beam, a longitudinal polarization measures its component along $\qhp$, and its component along $\nh$ is unobservable. For in-plane magnetization with $\vq\parallel\Mh$ the signal is longitudinal, $P_\para=2\eS(E/m)/[1+\eS^2(1+E^2/m^2)]$. It can be read with the native longitudinal polarization of a photocathode, but it carries a calculable background from the beam's spin magnetic moment, which adds $T/E$, or $6$--$16\%$ at 30--100~keV, to $P_\para$; the along-beam geometry avoids it. For $\vq\perp\Mh$ the convective dipole coupling vanishes, leaving exchange and the beam's spin current. The outgoing electrons are then polarized to $95\%$ along $\Mh$ at 100~keV, but at $6.6\times10^{-4}$ of the rate at allowed azimuths for $q=1$~\AA$^{-1}$, so that detection with an unpolarized beam requires a spin analyzer.

\begin{figure}[t]
  \centering
  \includegraphics[width=\columnwidth]{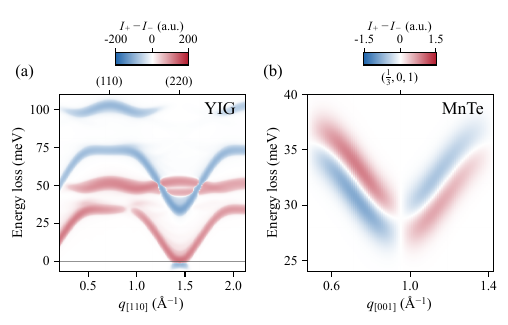}
  \caption{Kinematic spin-resolved magnon EELS in (a) YIG and (b) altermagnetic MnTe. The maps show the intensity difference $I_+-I_-$ for an unpolarized 30-keV incident beam, 10-nm specimens at 10~K, and ideal outgoing-spin analysis. The momentum cuts are $(H,H,0)$ for YIG and $(1/3,0,l)$ for MnTe. The beam is along $[001]$ for YIG and $-\bm b$, parallel to the N\'eel axis, for MnTe. The fixed transverse analyzer axes are $[110]$ for YIG and the direction of transverse momentum transfer at $(1/3,0,1)$ for MnTe. Gaussian energy and momentum resolutions are 3~meV and $0.02\,\mathrm{\AA}^{-1}$ FWHM. The YIG and MnTe color scales are symmetric logarithmic and linear, respectively. One a.u. corresponds to $10^{-13}$ probability per incident electron per meV in a $0.02\,\mathrm{\AA}^{-1}\times0.02\,\mathrm{\AA}^{-1}$ collection window.}
  \label{fig3}
\end{figure}

{To illustrate these spin-dependent signatures in specific materials, we calculate magnon EELS spectra of ferrimagnetic YIG and altermagnetic MnTe using an unpolarized incident beam and ideal transverse spin analysis. Figure~\ref{fig3} shows $I_+-I_-$, where $I_\pm$ are the intensities for outgoing spins parallel and antiparallel to the transverse analysis direction specified in the caption. This difference measures a transverse component of $\vP_f$ in Eq.~\eqref{eq:Pf}, weighted by the spin-summed intensity. Within the retained model, this transverse difference selects the dipole--exchange interference and vanishes when projectile exchange is omitted. The kinematic spectra exhibit opposite-sign contributions from modes of opposite handedness in the chosen scattering geometries. These material examples illustrate how projectile exchange makes magnon handedness accessible through spin-resolved electron spectroscopy.}

\textit{Discussion}---The exchange signature is a transverse polarization in the scattering plane that reverses with $\Mh$ and, for a polarized beam, a $\cos\phi$ asymmetry that also reverses with $\vP_b$. We confirm that no other process produces it by examining three competing processes, each of which either has the wrong symmetry or is too small. The first is spin-orbit (Mott) scattering~\cite{kessler1985}, the usual instrumental source of spin asymmetry. Its analyzing power lies along $\nh$, so it is orthogonal to the exchange signal, $90^\circ$ out of azimuthal phase with it, and even under $\Mh\ra-\Mh$. The second is the spin-dependent part of the beam current omitted from Eq.~\eqref{eq:a}~\cite{edstrom2016}, which lets the dipole channel polarize the beam on its own. Its relative amplitude is $q/2k$, so even with a phase $\delta$ from higher-order processes it gives a polarization of at most $(q/k)\sin\delta$, or $6\%$ of $P_\perp$ at 30~keV~\cite{companion}. The third is a Lorentz deflection of the whole beam by any in-plane component of the film's magnetic field, which produces an $\Mh$-odd left--right asymmetry. This asymmetry does not reverse with the beam polarization, so reversing the source polarization at fixed $\Mh$ removes it.

Multiple scattering changes the size of the signal but cannot create it. Elastic multiple scattering in the film~\cite{castellanos2025} multiplies $a$ and $\bm b$ by nearly the same phase and only rotates $\vP_\perp$ within the scattering plane. Bragg-scattered paths through $\vq'=\vq-\bm G$ carry dipole amplitudes along $\vK\times\vq'$ while the exchange amplitude keeps its direction, so their admixture can reduce $|\vP_\perp|$. A thin film and a momentum transfer away from strong reflections keep this admixture small.

%%% --->>> HARDWARE
Experimental implementation is feasible given current technological capabilities. In a conventional microscope the objective-lens field magnetizes the film along the beam axis, and a pixelated momentum-resolving camera, already used to observe magnons~\cite{kepaptsoglou2025,mm2025}, provides the azimuthal resolution that Eq.~\eqref{eq:cosphi} requires. For a lens field pointing along the beam's direction of travel, the moments align with the field, the ordered spins point against the beam, and $\vP_\perp$ lies along $-\qhp$, the direction of deflection (Fig.~\ref{fig1}). This experiment requires a beam angular spread below the scattering angle $q/k=11$~mrad at 30~keV, a condition easily met by the $2.25$~mrad used for magnon measurements in Ref.~\cite{kepaptsoglou2025}. A GaAs photocathode~\cite{pierce1976} with a strained superlattice~\cite{nishitani2005,maruyama2004} and a spin rotator can supply the transverse polarization; reversing the helicity of its laser reverses that polarization. Such sources deliver $P_b\approx0.85$~\cite{nishitani2005,maruyama2004}, and a 30~keV spin-polarized microscope beam has reached $82\%$~\cite{kuwahara2012,kuwahara2014}. Because the polarized beam acts as its own analyzer, no spin analyzer is needed after the sample, and because the signal is a ratio of counts taken with opposite helicities, beam current, film thickness, and the dipole cross section cancel. In the prefield of the objective lens the transverse beam spin precesses about the axis by about twice the image-rotation angle. This shifts the azimuthal origin of Eq.~\eqref{eq:cosphi}, which a fit of the $\cos\phi$ modulation determines, and leaves $P_b$ unchanged if the field is uniform across the beam. These components exist in separate instruments but have not yet been combined in one microscope.

The measurement favors low beam voltage. Because the dipole magnon rate is independent of beam energy, the counting time scales as $P_\perp^{-2}\propto\eS^{-2}$, which grows roughly as $T^3$, and the momentum transfer offers no compensation. For the acoustic magnon of YIG, circularly polarized about the net moment (net spin $5/2$ per formula unit)~\cite{cherepanov1993,princep2017}, the dipole probability per beam electron into a momentum window $\Delta^2q$ at transfer $q$ through a film of thickness $d$ is $p=2S_{\rm net}n\,F^2 r_e^2\,d\,\Delta^2q/q^2$~\cite{companion}, with $r_e$ the classical electron radius. For $\eS\ll1$, $P_\perp\simeq2\eS\propto q$ while $p\propto q^{-2}$, so the time depends on $q$ only through the magnetic form factor, which favors small $q$. For $q=1$~\AA$^{-1}$, $\Delta q=0.2$~\AA$^{-1}$, $d=10$~nm, and $P_b=0.85$, a $5\sigma$ measurement at 30~keV requires $2.5\times10^3$ magnon counts and takes
\be
t_{5\sigma}\simeq\frac{2\ {\rm h}}{f}\times\frac{1\ {\rm nA}}{I},
\label{eq:time}
\ee
where $I$ is the beam current and $f$ is the magnon fraction of counts in the energy window. Phonons in the window reduce $f$ but carry no $\Mh$-odd asymmetry. At 100~keV the time increases 38-fold, to $\simeq76~{\rm h}/f$ per nanoampere. Because no focused probe is needed, a parallel beam can deliver nanoamperes, far exceeding the picoampere current of the 1.3~nm probe in Ref.~\cite{kepaptsoglou2025}. The absolute rate depends on the magnetic structure factor, which for YIG at $1$~\AA$^{-1}$ lies several Brillouin zones out~\cite{companion}; $P_\perp$ does not. At 30~keV, where a spin-polarized microscope beam already exists~\cite{kuwahara2012}, the measurement takes hours rather than days.

{In conclusion, using magnon excitation as an example, we have shown that the exchange channel contributes to relativistic EELS, although its signature is suppressed in conventional unpolarized measurements. Spin-resolved detection or a transversely polarized source uncovers interference between the dipole and exchange channels, making it possible to measure the chirality of magnetic excitations. The predicted transverse polarization and left-right asymmetry provide clear experimentally measurable signatures of exchange effects, opening a new avenue for probing spin excitations in nanoscale materials via electron exchange in electron microscopes. Future developments in polarized electron sources, spin-resolved detection, and high-resolution EELS could further enhance the capabilities of electron microscopy in measuring spin-dependent quantum phenomena. Such measurements could provide deeper insight into the microscopic interactions and collective behavior underlying quantum magnetism.}

\bibliographystyle{apsrev4-1}
\bibliography{references}

%liuch.qf@outlook.com
\end{document}